\documentclass[11pt]{article}

\usepackage[utf8]{inputenc}
\usepackage[T1]{fontenc}
\usepackage{lmodern}

\usepackage[a4paper,margin=1in]{geometry}

\usepackage{amsmath}
\usepackage{amssymb}

\usepackage{graphicx}
\usepackage{subfig}
\usepackage{booktabs}
\usepackage{multirow}
\usepackage{placeins}

\usepackage[numbers]{natbib}

\usepackage[hidelinks]{hyperref}

\title{A Curated Literature Database for Monitoring More Than 30 Years of Ansys Granta Product Usage}

\author{
D. Mercier\thanks{Corresponding author: \texttt{david.mercier@synopsys.com}}\\
Ansys, part of Synopsys, ICME Team at Synopsys Innovation Group\\
France
}

\date{} 

\begin{document}
\maketitle
\begin{abstract}
Engineering and materials software is increasingly difficult to track in the scholarly and technical literature because publication volume is growing rapidly and software citation practices remain inconsistent. This is particularly true for the Ansys Granta product family, which is used for materials education, materials and process selection, sustainability-driven design, and enterprise materials information management. We present a structured and reproducible framework to consolidate evidence of \emph{operational} Granta usage and to support quantitative monitoring of adoption patterns, application domains, and technical impact. The framework is implemented as a curated reference database in \textit{Ansys Granta MI Enterprise}: bibliographic metadata are ingested semi-automatically (e.g., via DOI and citation-file parsing) and complemented by expert curation of usage descriptors (product, context, application domain, and technical depth), with relational links to authors and institutions. Downstream analytics are performed with Python, dashboards, and bibliometric/network visualization tools to enable reproducible querying and reporting. As of September~2025, the database contains more than 1{,}100 curated records spanning journals, conferences, theses, books, patents, standards, and reports, and supports rapid retrieval of validated case studies, reproducible literature reviews, and technology scouting. Example analyses highlight dominant domains, key institutions, and recurring integrations with CAD/CAE/FEM environments. Overall, the approach converts heterogeneous software-usage evidence into structured, analyzable knowledge to improve visibility of engineering software impact and to support evidence-based assessment and strategic decision-making.
\end{abstract}

\vspace{0.75em}

\section*{Graphical Abstract}

\begin{center}
\includegraphics[width=\linewidth]{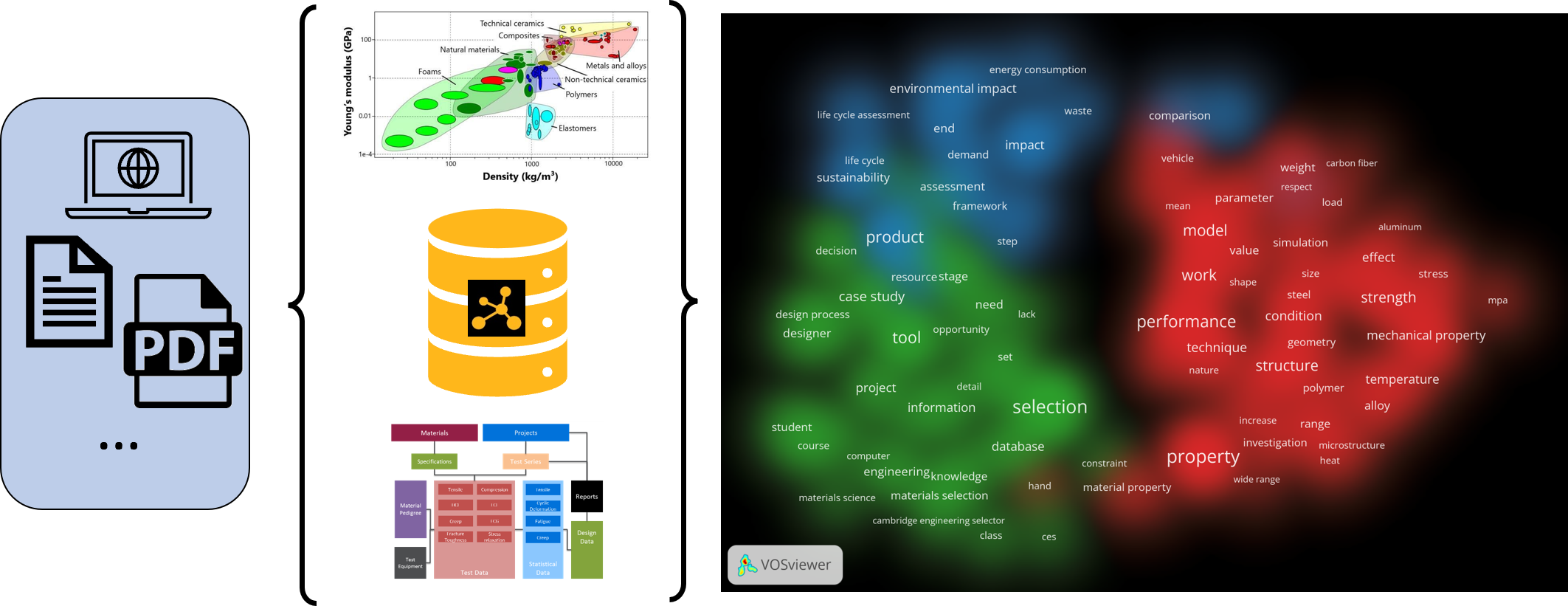}
\end{center}

\section{Introduction}
Software tools and computational codes have become indispensable to contemporary science, engineering, and industry. Beyond their role as numerical engines for simulation and data analysis, software products increasingly act as vehicles for methodological innovation, repositories of domain-specific knowledge, and enablers of scientific reproducibility \cite{johanson_software_2018, arvanitou_software_2021}. In many disciplines, the adoption of a software solution shapes research practices, constrains which classes of problems can be addressed, and influences how results are validated and disseminated. For software developers and technology providers, understanding how products are used, cited, and embedded in the scientific and technical literature is therefore strategically important, informing product development, customer engagement, and long-term research and innovation planning~\cite{muhammad_ansys_2020, masuadi_trends_2021, tomaszewski_analyzing_2023}.

From an analytical perspective, bibliometrics offers a mature toolkit to characterize diffusion and impact~\cite{pan_examining_2018, dervis_bibliometric_2020, boukhlif_decade_2023}. While traditionally applied to authors, institutions, and topics, these methods must be adapted to treat software products as first-class objects of scholarly communication. Co-authorship and co-citation networks reveal communities of practice; keyword co-occurrence and thematic mapping expose application structures and emerging themes; and temporal indicators reveal adoption dynamics. Crucially, when bibliometric indicators are combined with curated metadata describing \emph{how} a tool is used (e.g., as a materials selection environment, a data management backbone, or a workflow component coupled to CAD/CAE), the resulting analysis becomes directly actionable for both scientific assessment and technology strategy.

Software-centric bibliometrics faces several fundamental challenges~\cite{donthu_how_2021}. Software citation practices remain highly heterogeneous: authors may cite a software name, a methodological or companion paper, a user manual, a versioned release associated with a Digital Object Identifier (DOI), or simply mention the tool in the text without providing any formal reference. In addition, software evolves through successive releases, forks, and integrations, making static bibliographic records poorly suited to capturing longitudinal adoption patterns and usage dynamics. These difficulties are compounded by the fact that evidence of software usage is scattered across heterogeneous document types—including journal articles, conference proceedings, theses, technical reports, books, patents, and standards—each following distinct citation conventions and exhibiting highly variable metadata quality. As a result, even large-scale open bibliographic infrastructures struggle to reliably identify and track software usage. For example, in OpenAlex, software is not treated as a first-class scholarly entity, and explicit software usage fields are generally absent from publication records \cite{priem_openalex_2022}. Consequently, identifying publications that operationally use a given software product often requires full-text inspection or heuristic keyword searches, which are error-prone and difficult to scale. Together, these factors substantially limit the effectiveness of generic bibliographic databases for systematic, software-oriented monitoring and motivate the need for dedicated, curated approaches.

These difficulties are particularly acute for the Ansys Granta product family \cite{AnsysMaterials2025}, whose tools are used across a wide range of scientific and engineering contexts, including materials science and modelling~\cite{kohler_clinching_2020, deepak_data_2025}, product design~\cite{delibas_thermo-economic_2025, musenich_dhat_2025}, sustainability~\cite{nicolalde_selection_2022, hamade_life_2020}, data management~\cite{barreneche_new_2015, perdu_estor_2018}, and Integrated Computational Materials Engineering (ICME) and materials informatics~\cite{arnold_information_2016, marsden_2016, campagna_streamlining_2025}. In addition to research and industrial applications, Granta tools are widely used in engineering education, further increasing the diversity of usage contexts and publication venues. In such settings, ad hoc literature searches and informal reference lists quickly become inadequate: they lack traceability, consistent metadata, relational structure, and scalability.

To address this need, we developed a curated reference database implemented in \textbf{Ansys Granta MI Enterprise}~\cite{AnsysGrantaMI2025}, dedicated to the systematic identification and characterization of documents reporting the \emph{operational} use of Ansys Granta products. The approach is aligned with prior work on curated software-use evidence, engineering knowledge capture, and structured data frameworks~\cite{cebon_engineering_2006, dong_transformation_2021, hearley_robust_2023}. The objectives of this work are threefold: (i) to define a relational database architecture and data model tailored to software-centric literature monitoring; (ii) to establish a semi-automated ingestion and expert-curation workflow that explicitly distinguishes intrinsic bibliographic metadata from extrinsic software-usage descriptors; and (iii) to demonstrate a set of reproducible bibliometric and visualization analyses that transform curated records into quantitative indicators and interactive dashboards.

The remainder of this paper is organized as follows. Section~\ref{sec:database} presents the Granta product family and describes the database architecture and curation workflow. Section~\ref{sec:analytics} introduces the bibliometric and analytical framework used to derive indicators and networks. Section~\ref{sec:results} reports results on temporal adoption, key institutions and collaboration structures, disciplinary positioning, material-family usage, and integration with CAD/CAE/FEM tools.

\section{Ansys Granta Products, Database Architecture, and Curation Workflow}
\label{sec:database}

This section introduces the Ansys Granta product family and describes how the reference database is implemented in Ansys Granta MI Enterprise. It details the database architecture, the data model, and the semi-automated workflow used to collect, curate, and enrich literature records explicitly referencing Ansys Granta products \cite{AnsysMaterials2025}.

\subsection{Ansys Granta products}

The Ansys Granta product family comprises a portfolio of software solutions dedicated to materials information management, materials selection, and data-driven materials engineering. It supports activities ranging from engineering education and early-stage design to industrial data management and ICME workflows. The foundations of these tools were laid by the materials selection methodology developed in the 1980s and early 1990s by M.~F.~Ashby and D.~Cebon, which introduced systematic, property-based approaches to materials choice and trade-off analysis, notably through the use of materials property charts (commonly referred to as "Ashby plots"). These concepts were formalized in seminal publications and rapidly adopted within engineering design and education \cite{cebon_materials_1994,ashby_materials_1999, cebon_engineering_2006}. Building on this methodological framework, Granta Design Ltd. was established in Cambridge in 1992 to translate these academic principles into industrial and educational software tools, leading to the development of the Granta product line for materials databases, selection, and information management. In 2019, Granta Design Ltd. was acquired by Ansys Inc., and the Granta products became an integral part of the Ansys software ecosystem, strengthening their coupling with simulation-driven engineering, digital materials, and ICME workflows \cite{AnsysMaterials2025}. Since 2026, the Granta product family has been developed and maintained within Synopsys, further extending its integration into electronic design automation and semiconductor-oriented engineering platforms \cite{SynopsysHomepage2026}.

\textbf{Granta MI Enterprise} (often referred to simply as \textbf{Granta MI}) is the core enterprise materials information management system developed by {Ansys} \cite{AnsysGrantaMI2025}. It provides a centralized environment for storing, curating, and disseminating materials, process, and simulation data, together with associated technical knowledge, across organizations. The platform supports structured data models, access control, versioning, and full traceability, and integrates with external CAD/CAE and simulation workflows. Granta MI is widely deployed in industrial and research settings to ensure consistency, reliability, and governance of materials data throughout the product development lifecycle. Furthermore, the system aligns with the FAIR data principles, promoting materials information that is findable, accessible, interoperable, and reusable, thereby facilitating transparency, long-term reuse, and cross-platform data exchange \cite{wilkinson_fair_2016}.

\textbf{Granta EduPack} is a materials selection and education-oriented software environment broadly adopted in engineering teaching and academic research \cite{AnsysGrantaEduPack2025}. It provides curated databases of materials and processes, interactive selection charts, performance indices, and pedagogical resources that support systematic materials selection, eco-design, and sustainability-driven design. EduPack is extensively used in undergraduate and graduate curricula, as well as in early-stage conceptual and comparative design studies reported in the literature.

\textbf{Granta Selector} is a professional decision-support tool primarily targeted at industrial engineering and product development contexts \cite{AnsysGrantaSelector2025}. It extends the materials selection methodology to large-scale industrial databases, enabling the systematic screening and comparison of candidate materials and processes under combined technical, economic, and environmental constraints.

Both EduPack and Selector incorporate specialized analytical modules, notably the \textit{Eco Audit} and \textit{Synthesizer} tools. The Eco Audit module supports streamlined life-cycle thinking by providing simplified life cycle analysis (LCA) data and environmental performance indicators, enabling rapid estimation and comparison of embodied energy, carbon footprint, and other impacts during early design stages. The Synthesizer module enables the exploration of hypothetical or hybrid materials (e.g., composites), estimating effective properties and costs to help identify gaps in the materials property space and compare advanced material solutions with conventional engineering alternatives. Together, these tools facilitate informed, sustainability-aware, and innovation-oriented design decisions.

Together, these products form a coherent ecosystem that supports materials data management, materials selection, and data-driven engineering across education, research, and industry, underpinned by shared access to core materials property datasets that can be extended with advanced materials, process, supplier, test, and standards-based data to inform design, selection, qualification, and simulation workflows~\cite{AnsysMaterialsDataLibrary2025}. Prior to the acquisition of Granta Design Ltd.\ by Ansys, earlier generations of these tools were marketed under the Cambridge Engineering Selector (CES) brand, notably \textit{CES EduPack} and \textit{CES Selector}; these legacy names still appear in portions of the historical literature and are therefore considered during corpus identification and curation.

It is important to note that Ansys Granta products should not be interpreted or compared in bibliometric terms in the same way as general-purpose numerical or scientific software such as finite-element solvers, electronic-structure codes, or mathematical computing environments. Tools such as VASP or MATLAB are typically cited as primary computational engines that directly generate numerical results, often appearing systematically in methods sections and reference lists. By contrast, Granta products occupy a more specialized and enabling position within engineering workflows: they provide curated materials, process, and sustainability data, along with selection and decision-support methodologies, that inform design choices, parameterization, and interpretation rather than acting as standalone solvers. As a result, their usage is frequently implicit, embedded upstream of simulation or design activities, and not always explicitly documented in formal citations. This niche positioning explains both the heterogeneity of reporting practices observed in the corpus and the need for expert curation to capture meaningful evidence of Granta product usage in the literature.

\subsection{Database overview and record structure}

\begin{figure}[h]
\centering
\includegraphics[width=0.95\textwidth]{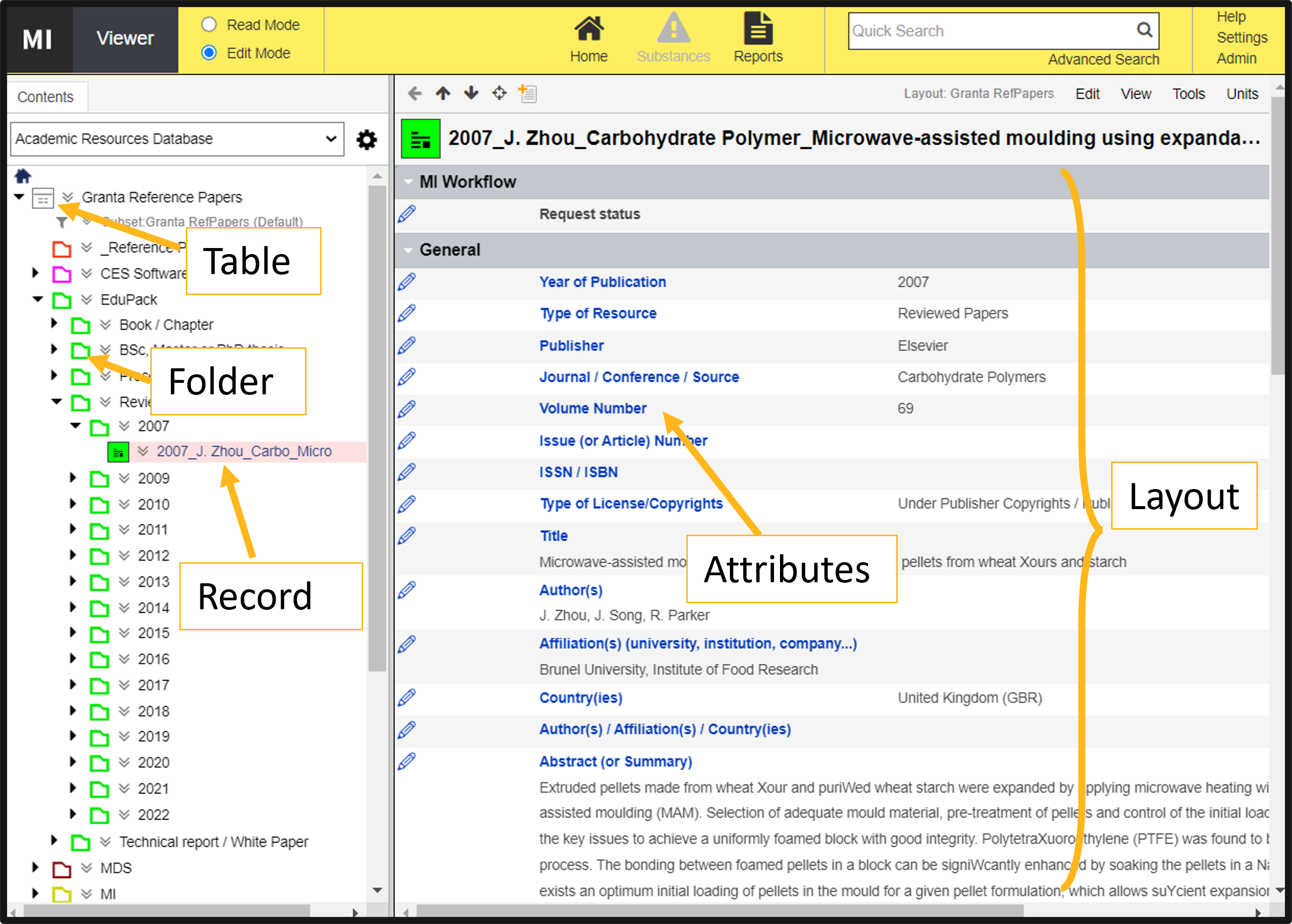}
\caption{Example of the Ansys Granta MI Viewer web interface illustrating the notions of table, record, attribute, and layout. Attributes are displayed in light blue to indicate that they are interactive: clicking on an attribute name opens a dedicated page providing its detailed definition and metadata.}
\label{fig:MIViewerDetailled}
\end{figure}

The reference database is implemented in Ansys Granta MI Enterprise itself, which provides a structured, schema-driven environment for storing, curating, and querying heterogeneous technical information. Granta MI combines a relational database backend with a domain-specific data model optimized for materials, engineering data, and technical documentation~\cite{Cebon_Ashby_2006, Ren_2011, allec_case_2024}. Records are stored in typed tables, linked through controlled relationships, and accessed through Application Programming Interfaces (APIs) and web interfaces, enabling both interactive exploration and automated analysis. One of the web-based user interfaces of Granta MI, the \textit{MI Viewer}, is illustrated in Fig.~\ref{fig:MIViewerDetailled}.

It is important to note that Granta MI is not intended to function as a conventional reference management tool in the sense of widely used solutions such as Zotero or Mendeley, which focus on citation handling, PDF management, and manuscript-centric workflows~\cite{zhang_comparison_2012}. While Granta MI provides by default a subset of comparable capabilities for storing and organizing bibliographic metadata, its primary value lies elsewhere: namely, in its ability to natively link literature references with structured materials data, engineering properties, and reference materials databases within a unified information model. This tight coupling between publications and underlying materials data represents a key advantage when the objective is to contextualize scientific papers with quantitative materials information and to support data-driven analyses beyond traditional reference management.

\begin{figure}[h]
\centering
\includegraphics[width=0.95\textwidth]{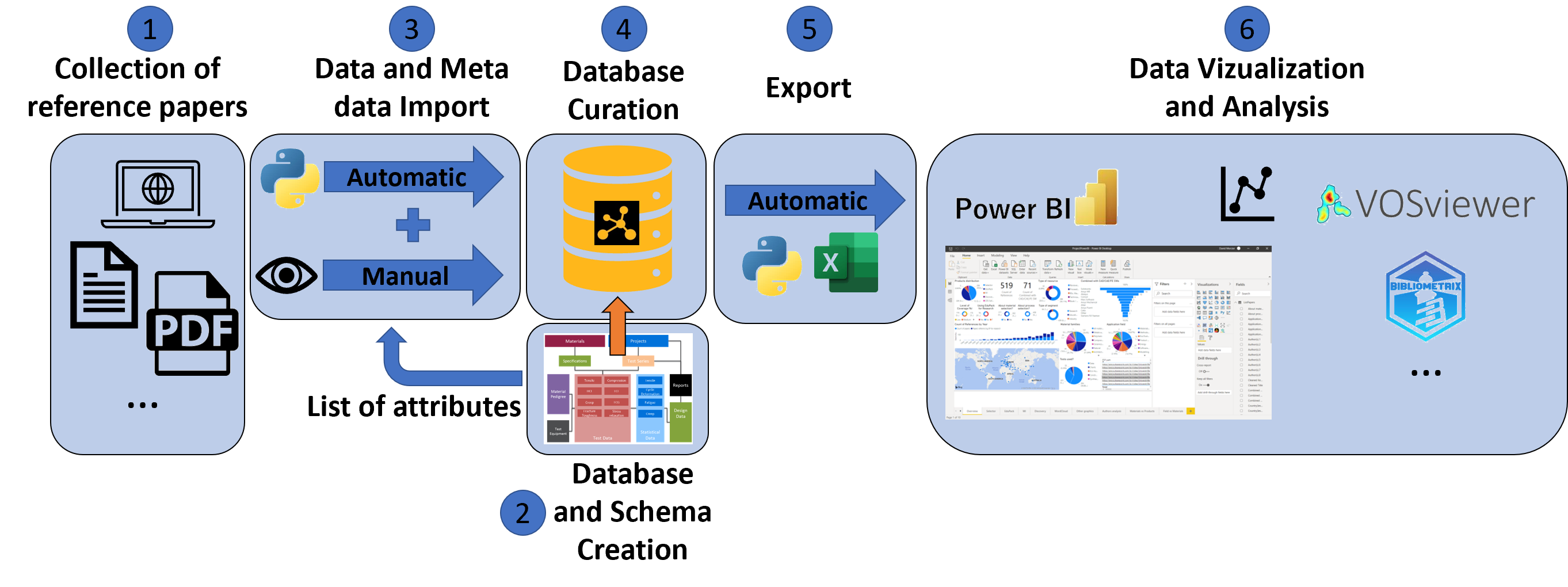}
\caption{Semi-automated curation and analytics workflow used to populate and exploit the Granta MI reference database. Intrinsic bibliographic metadata are extracted automatically from publisher web pages and citation files, whereas extrinsic usage and application descriptors are obtained through manual expert analysis, curated through quality-control procedures, and exported for bibliometric, network, and visual analytics.}
\label{fig:MIWorkflow}
\end{figure}

Within this framework, the literature reference database is organized as a set of interlinked tables that together form a \textit{technical knowledge map} of Granta product usage in scientific and engineering publications. The core entity is the document record, representing a single publication. The corpus assembled for this study comprises more than 1{,}100 documents, including journal articles, conference papers, theses, standards, and patents, all written in English (Step~1 of the workflow, Fig.~\ref{fig:MIWorkflow}). Each document is uniquely identified (e.g.\ \textit{Record Name}, \textit{Short Name}) and associated with bibliographic, contextual, and technical metadata.

\begin{figure}[h]
\centering
\includegraphics[width=0.5\textwidth]{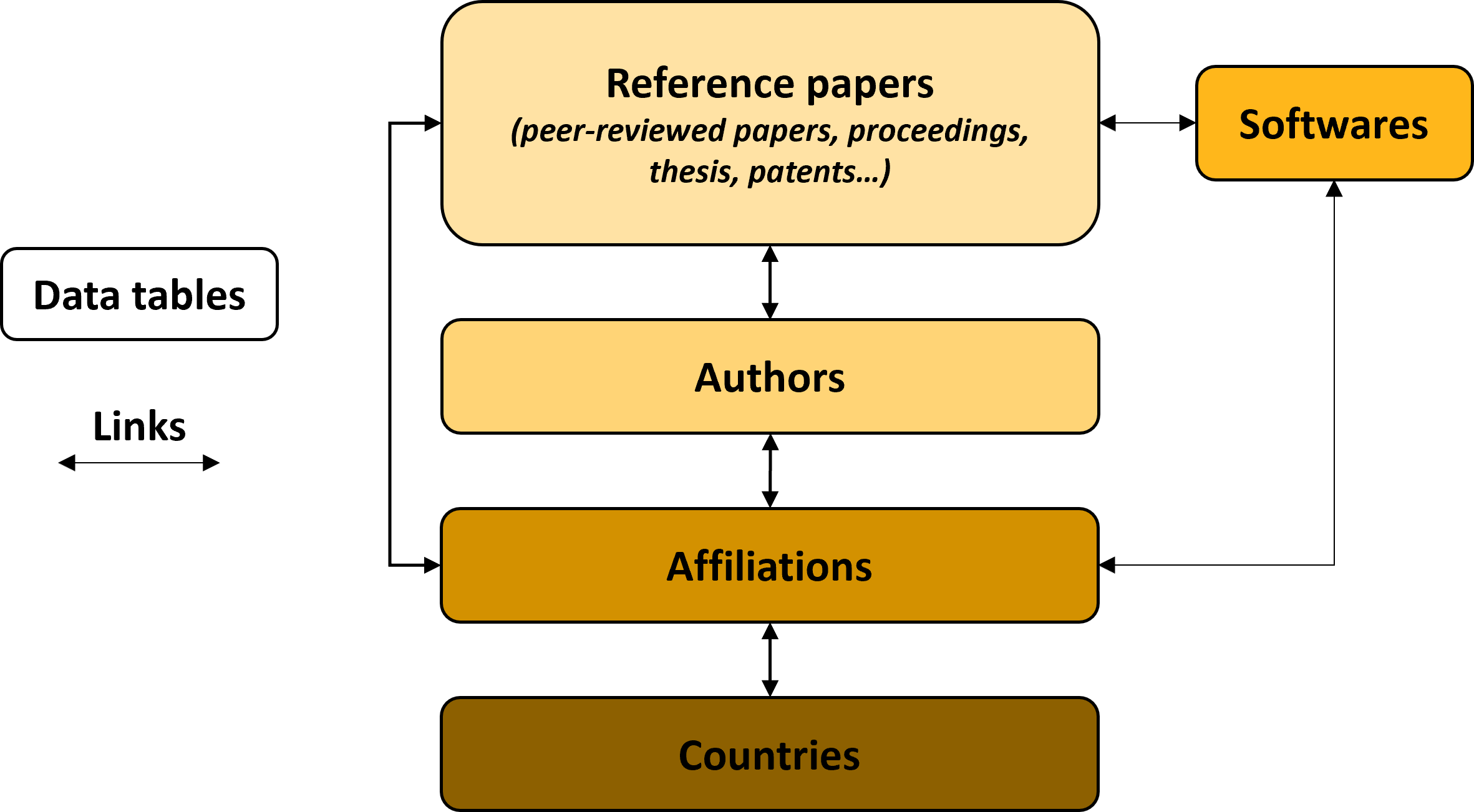}
\caption{Illustrative representation of the Granta MI database schema used for the literature reference database. Core document records are linked to relational tables describing authors, affiliations, countries, products, and application domains. Double-arrow links indicate bidirectional navigability between tables, allowing users to jump interactively from one record (e.g., a paper) to related records (e.g., its affiliations or authors) by clicking on the corresponding links.}
\label{fig:databaseSchema}
\end{figure}

A typical document record, illustrated by the example shown in Fig.~\ref{fig:MIViewerDetailled} (right panel), follows a structured layout organized into thematic sections that capture bibliographic information, product usage, application context, and traceability metadata. The underlying data model and relational structure are defined during the schema design phase (Step~2 of Fig.~\ref{fig:MIWorkflow}) and are summarized in Fig.~\ref{fig:databaseSchema}.

From a conceptual standpoint, the metadata associated with each record are divided into two complementary categories: \textit{intrinsic} and \textit{extrinsic} properties. Intrinsic properties correspond to information inherent to the publication itself and can be obtained from bibliographic sources, whereas extrinsic properties describe how the document relates to Ansys Granta technologies and to its technical and application context, and therefore require expert interpretation.

\paragraph{Intrinsic properties: bibliographic metadata.}
This section captures the intrinsic identity and provenance of the publication, including:
\begin{itemize}
  \item bibliographic descriptors such as title, year of publication, document type, journal or conference, volume, and issue;
  \item authors, affiliations, and countries;
  \item publisher and copyright information;
  \item abstract or summary, keywords, and language;
  \item persistent identifiers and reference information (e.g.\ DOI and cited references).
\end{itemize}
These fields ensure unambiguous identification of each document and support citation tracking and bibliographic consistency. Intrinsic metadata are primarily acquired automatically during the ingestion phase using scripted workflows that rely on standard bibliographic sources such as digital object identifiers and citation records (e.g.\ \texttt{.bib}, \texttt{.ris}). The resulting fields provide a consistent and reproducible bibliographic description of each reference. Higher-level intrinsic descriptors (e.g.\ keywords or language) may be completed, verified, or corrected manually when necessary to ensure accuracy and consistency.

\paragraph{Extrinsic properties: Granta product usage.}
Extrinsic metadata describe how a publication relates to the Granta software ecosystem and cannot be reliably inferred from bibliographic sources alone. This section captures how Ansys Granta technologies are involved in each publication, including:
\begin{itemize}
  \item the principal Granta product used (e.g.\ \textit{EduPack}, \textit{Selector});
  \item product version and optional sub-products;
  \item the usage context (academic research, education, benchmarking, competitive analysis);
  \item potential coupling with external CAD, CAE, and FEM software tools.
\end{itemize}
These descriptors establish explicit links between publications and specific elements of the Granta product portfolio, enabling systematic analysis of product adoption, integration patterns, and technological diffusion.

\paragraph{Extrinsic properties: application domain and technical context.}
This section captures the technical scope and intent of each publication through curated extrinsic descriptors, including:
\begin{itemize}
  \item application domains and disciplinary areas (e.g., product development, mechanical or structural engineering, sustainability, etc.); these application fields were mapped to the Organisation for Economic Co-operation and Development (OECD) \emph{Fields of Science and Technology} (FOS, 2007) classification \cite{Grabowska_2019};
  \item user research segment (academia, industrial, or education);
  \item breadth, depth, and overall scope of product utilization;
  \item description of the relevant material classes and manufacturing process families considered;
  \item level of use of Ansys Granta tools and functionalities (e.g., databases, charts, performance indices, Eco Audit, Synthesizer, etc.);
  \item classification indicators identifying whether the study addresses materials selection or process selection;
  \item supplementary notes, observations, and comments.
\end{itemize}

Together, these fields provide a structured representation of both \textit{what the document is technically about} and \textit{how Granta tools are used within the reported work}. Because such information is rarely encoded explicitly in bibliographic metadata, these extrinsic properties are derived primarily through expert reading and interpretation.

All extrinsic metadata describing product usage are obtained primarily through manual expert analysis. Multiple curators (mostly database users) examine documents to identify the Granta products involved, the level and mode of usage, and the associated technical domains. Whenever possible, these annotations are discussed and cross-validated to improve consistency and reduce subjective bias.

\paragraph{Location and links.}
This section consolidates traceability information, including:
\begin{itemize}
  \item persistent identifiers and external access information (DOI and/or URL);
  \item internal document paths (e.g.\ local PDF storage) for long-term retrieval;
  \item relational links within Granta MI to associated entities (authors, institutions, products, sources, and application domains).
\end{itemize}
Together, these elements support systematic cross-referencing across the database and ensure that each record remains verifiable and reusable, while enabling both tabular and graph-based exploration of relational structures.

\subsection{Relational navigation and link visualization}
\label{subsec:linkviz}

Beyond tabular exploration and attribute-based filtering, Granta MI provides dedicated tools for visual exploration of the relational structure of the database. In particular, the \textit{Links Visualizer} interface enables interactive inspection and navigation of the network of relationships between records stored across different tables, including documents, authors, institutions, products, and application domains.

\begin{figure}[h]
\centering
\includegraphics[width=0.75\textwidth]{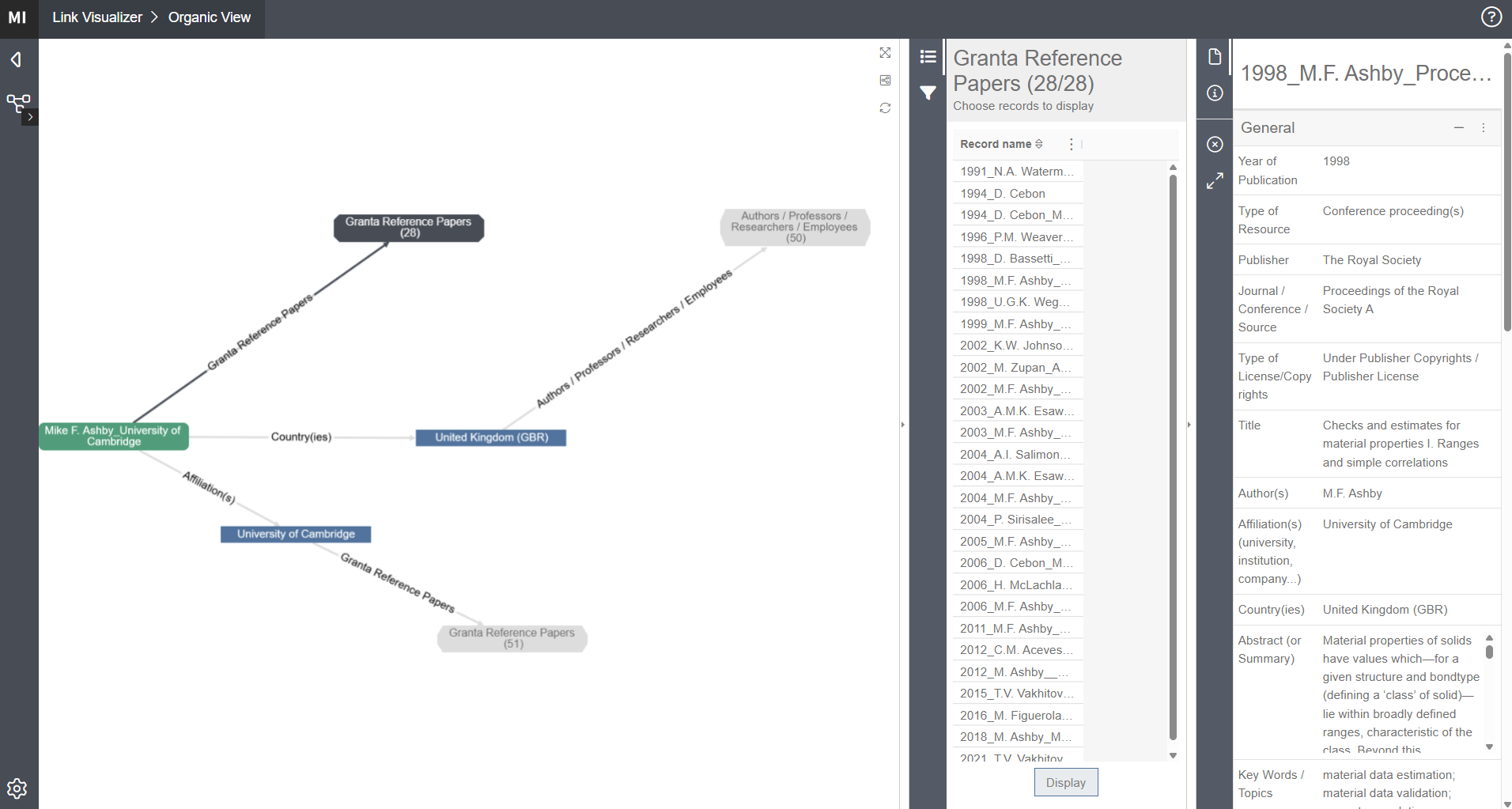}
\caption{Example of the Links Visualizer tool in the Ansys Granta MI Viewer web interface, highlighting connections between records in different tables. The example shows the network of links associated with the author M.~F.~Ashby, including related documents, affiliations, and countries.}
\label{fig:linksViz}
\end{figure}

As illustrated in Fig.~\ref{fig:linksViz}, selecting a given entity (e.g.\ an author, an institution, or a product) dynamically reveals its direct and indirect connections to other records, allowing users to rapidly identify collaboration patterns, product dissemination pathways, and cross-domain linkages. This graph-based representation complements the tabular interfaces of MI Viewer and MI Explore by providing a structural and intuitive view of database connectivity that is difficult to infer from tables alone.

These relational visualization capabilities are particularly valuable for exploratory analysis and data quality control, as they facilitate the detection of missing links, inconsistent associations, or unexpected connectivity patterns within the curated data. More broadly, they contribute to transforming the reference database from a static collection of documents into a navigable \textit{knowledge graph} of Granta product usage in the scientific and engineering literature, thereby supporting both qualitative insight generation and downstream network-based analytical workflows.

\begin{figure}[h]
\centering
\includegraphics[width=0.95\textwidth]{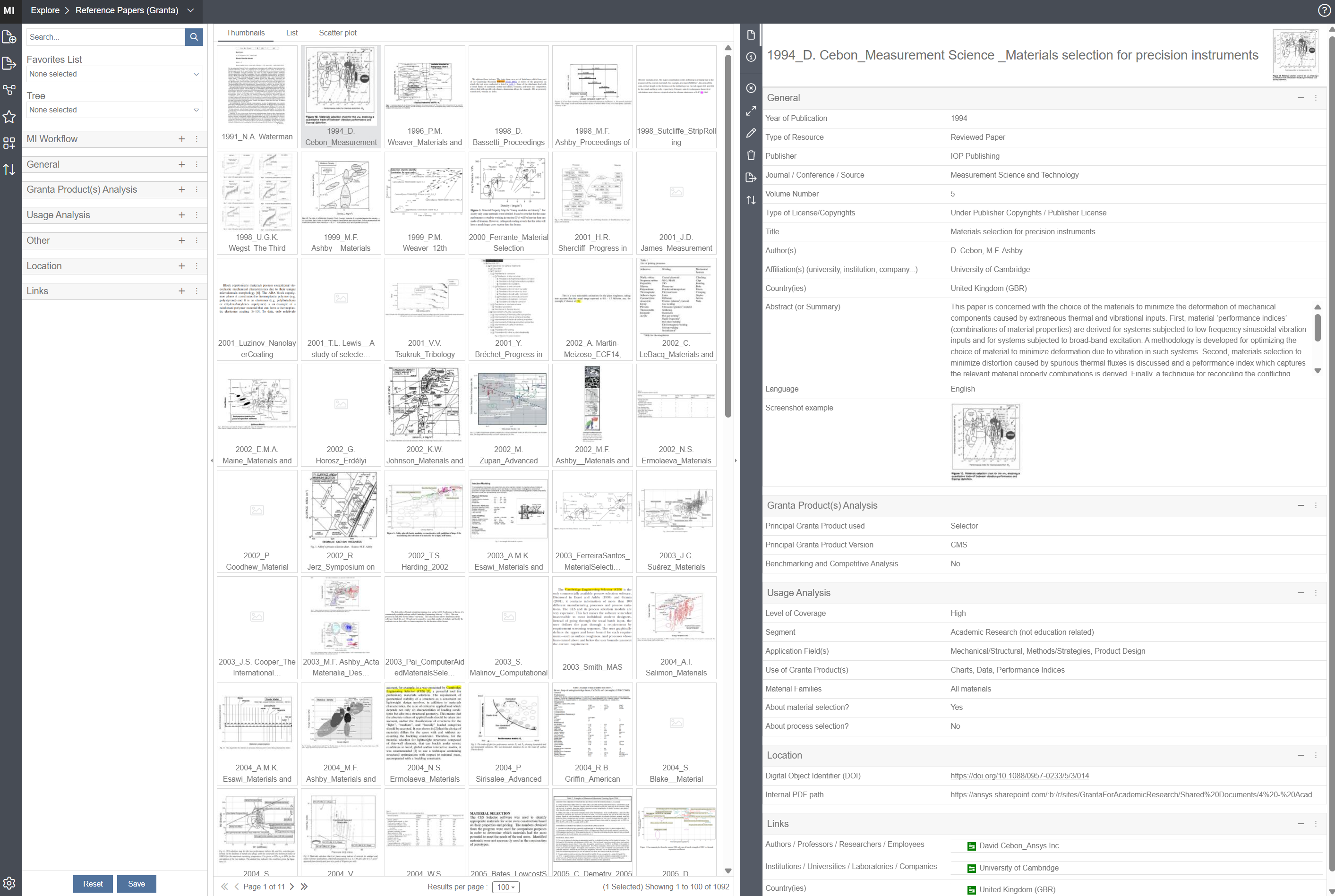}
\caption{Example of the Ansys Granta MI Explore web interface. The left panel provides filtering and search tools to select subsets of documents or locate specific references within the corpus, while the right panel displays a curated document record for the paper \textit{``Materials selection for precision instruments''} by D.~Cebon and M.~F.~Ashby~\cite{cebon_materials_1994}, one of the earliest publications referring to a Granta product.}
\label{fig:MIExplore}
\end{figure}

\FloatBarrier

\FloatBarrier

\FloatBarrier

\subsection{Curation and analytics workflow}

A fourth step consists of database curation and quality control, combining manual review by database users and domain experts with automated consistency checks implemented in Python to identify missing, incomplete, or inconsistent values. In a fifth step, the curated data are exported using dedicated Python scripts or Granta MI export functions to produce structured tabular datasets (e.g.\ Python data frames serialized in \texttt{.json} format or \texttt{.xls} files). These datasets are then used in a sixth step for downstream bibliometric, network, and visual analytics employing established tools~\cite{Cobo_2011}, including Microsoft Power BI~\cite{Li_2025}, VOSviewer~\cite{van_eck_software_2010}, and the \texttt{bibliometrix} libraries for Python and R~\cite{Dervis2020Bibliometric}.

To support interactive exploration and reporting, the analytical results are integrated into a set of visual dashboards implemented using Microsoft Power BI. These dashboards provide dynamic views of publication trends, product usage distributions, collaboration networks, and thematic structures, with filtering capabilities by time period, product, application domain, document type, and institution. Additional network visualizations are generated using VOSviewer and the \texttt{bibliometrix} libraries to produce co-authorship, co-citation, and keyword maps. These visual representations complement the quantitative indicators and facilitate qualitative interpretation of usage patterns, community structures, and emerging trends.

Together, these components define a \textbf{relational and analytical representation} in which intrinsic bibliographic properties and extrinsic usage descriptors are systematically linked to products, application domains, authors, institutions, and sources. As illustrated in Figs.~\ref{fig:MIViewerDetailled}, \ref{fig:MIExplore}, and~\ref{fig:databaseSchema}, this structured organization allows the database to function not merely as a document repository, but as a \textit{queryable, multi-dimensional map of how Granta technologies are used in the scientific and engineering literature}, supporting large-scale querying, cross-referencing, and downstream bibliometric, network, and visualization workflows.

\FloatBarrier

\section{Bibliometric and Analytical Framework}
\label{sec:analytics}

This section presents the bibliometric and analytical framework used to transform the curated reference database into quantitative indicators, network representations, and interactive visual summaries of Ansys Granta product usage.

\subsection{Corpus definition}

The bibliometric corpus is derived from the curated reference database described in Section~\ref{sec:database}. As of September~2025, it comprises more than 1{,}100 records collected from peer-reviewed journal articles, conference proceedings, books, theses, technical reports, standards, and patents that explicitly report the use of one or more Ansys Granta products. The dataset spans publications from the early 1990s to 2025, reflecting more than three decades of software adoption and usage. As a complementary and coarse indicator of broader visibility beyond the curated and manually validated corpus, simple keyword searches performed on Google Scholar (December~2025) return on the order of hundreds to thousands of results for the main product names, including more than 1{,}500 hits for \textit{``Granta EduPack''}, approximately 410 for \textit{``Granta CES''}, 370 for \textit{``Granta MI''}, and 220 for \textit{``Granta Selector''} \cite{GoogleScholar2025}. Although such counts are inherently approximate and depend on query formulation and indexing policies, they nonetheless suggest a substantial and sustained presence of Granta products across the wider scientific and technical literature.

A key characteristic of this corpus is that it is restricted to documents in which Granta products are used in a \emph{substantive and operational manner}. In particular, records were retained only when the software is employed to:
(i) access, curate, or manage materials and process data,
(ii) support materials selection or screening workflows (e.g.\ using "Ashby charts"),
(iii) generate comparative charts, property maps, or trade-off analyses, or
(iv) integrate materials data into engineering, design, or simulation workflows.
Publications merely citing Granta products as illustrative examples or in background or state-of-the-art sections, without actual usage in the reported methodology or results, were systematically excluded.

Each retained record is associated with structured metadata describing bibliographic attributes (year, venue, authors, affiliations, country), product usage (product name, version when available, and usage context), application domains, and curated technical descriptors. The corpus includes documents from academic research, industrial applications, and engineering education, and spans multiple scientific and engineering fields, including materials science, mechanical design, sustainability, data management, and ICME.

For the analyses reported in this paper, records were filtered to retain only documents with complete bibliographic information and at least one explicit and operational reference to a Granta product. Duplicates, incomplete entries, and non-relevant mentions were removed during the quality-control phase. The resulting dataset therefore constitutes a \emph{software-centric} bibliographic corpus designed to support both descriptive and network-based analyses of product usage patterns rather than generic citation analysis.

\subsection{Bibliometric indicators}

\begin{figure}[h]
\centering
\includegraphics[width=0.65\textwidth]{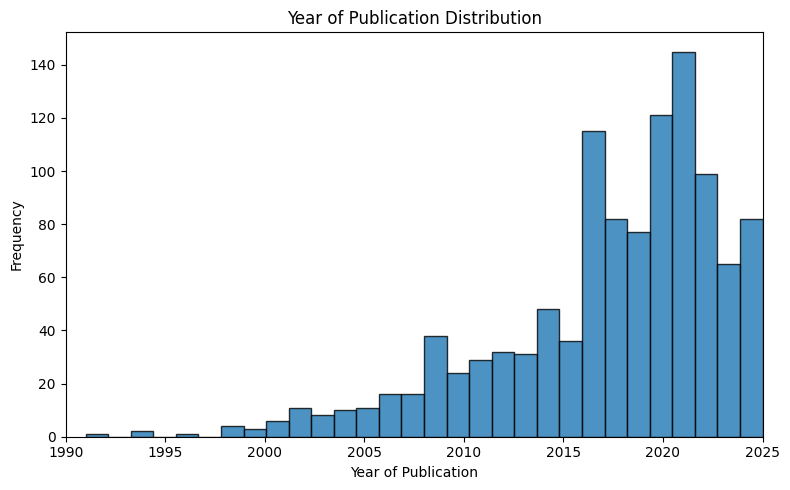}
\caption{Temporal distribution of publications in the curated corpus by year of publication (1990--2025), illustrating the long-term evolution and recent growth of Ansys Granta product usage in the scientific and technical literature.}
\label{fig:Paper_Year}
\end{figure}

\begin{figure}[h]
\centering
\includegraphics[width=0.95\textwidth]{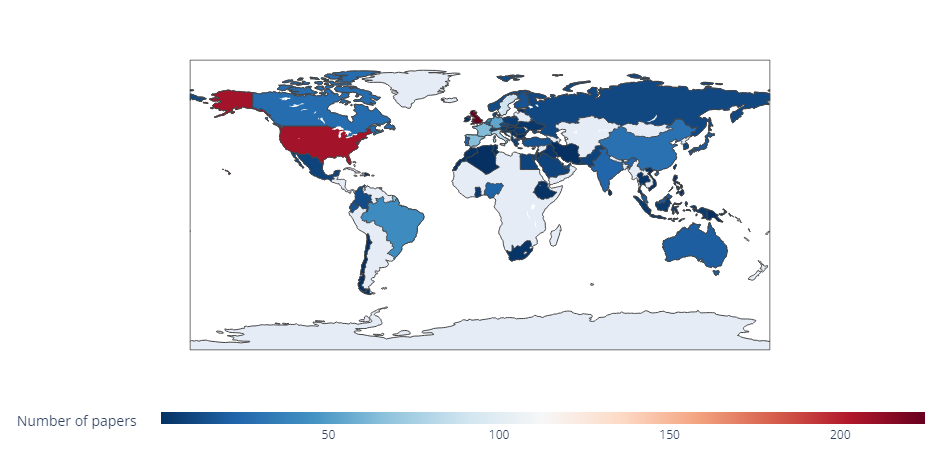}
\caption{Geographical distribution of publications in the curated corpus based on the country information extracted from author affiliations.}
\label{fig:PapersCountry}
\end{figure}

A first level of analysis relies on standard descriptive bibliometric indicators to characterize the size, growth, and distribution of the corpus. These indicators include annual publication counts and cumulative publication curves (Fig.~\ref{fig:Paper_Year}), as well as distributions of documents by document type (Table~\ref{tab:resource_type}), country (Fig.~\ref{fig:PapersCountry}), product, application domain, and usage segment.

\FloatBarrier

The geographical distribution of publications, inferred from author affiliation countries, indicates that the corpus is dominated by documents originating from the USA, the UK, and Europe more broadly (Fig.~\ref{fig:PapersCountry}). In terms of document types, the collection is primarily composed of peer-reviewed journal articles, complemented by conference proceedings, theses, and a smaller number of technical reports and standards/patents categories (Table~\ref{tab:resource_type}).

\begin{table}[h]
\centering
\caption{Distribution of records by document type in the curated corpus.}
\label{tab:resource_type}
\begin{tabular}{l r}
\toprule
Resource Type & Count \\
\midrule
Reviewed Paper & 596 \\
Conference proceeding(s) & 219 \\
BSc, Master or PhD Thesis & 163 \\
Technical Report / White Paper & 88 \\
Standards / Patents & 47 \\
\bottomrule
\end{tabular}
\end{table}

\FloatBarrier

\FloatBarrier

Additional descriptive indicators were derived from curated metadata, such as the frequency of product mentions, the distribution of usage contexts (research, industry, education), and the relative contribution of publication venues and scientific domains. Together, these indicators provide a quantitative overview of the diffusion and evolution of Ansys Granta technologies across the scientific and technical literature.

\subsection{Author collaboration structure}
\label{subsec:authors_network}

More than 3{,}300 distinct authors were identified across the curated corpus, reflecting the broad and multidisciplinary user base of Ansys Granta products. To focus on recurring and structurally meaningful collaborations, the co-authorship analysis was restricted to authors meeting a minimum document threshold of three publications. This filtering yields a reduced network composed primarily of stable collaboration groups, typically forming clusters of at least ten authors.

\begin{figure}[h]
\centering
\includegraphics[width=0.95\textwidth]{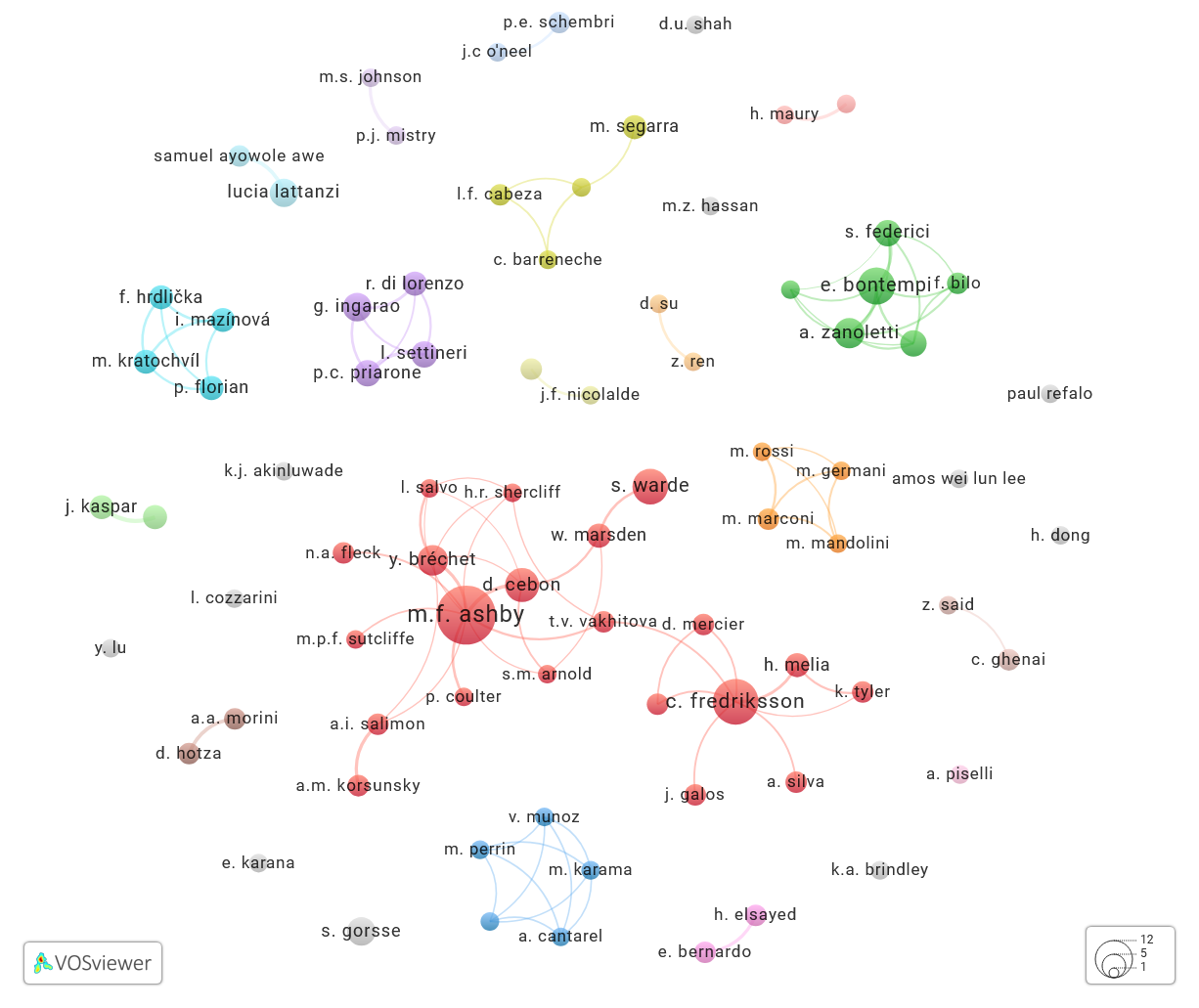}
\caption{Co-authorship network generated using VOSviewer (overlay visualization). Nodes represent authors and links indicate co-authored publications; node size is proportional to the number of documents and link strength reflects collaboration intensity. The overlay color encodes the average publication year, revealing both historically rooted author communities (early average years) and more recent clusters (late average years).}
\label{fig:AuthorsNetwork}
\end{figure}

A co-authorship network was generated using VOSviewer from this filtered set. In Fig.~\ref{fig:AuthorsNetwork}, nodes represent authors and links denote co-authored publications, while link strength reflects the intensity of repeated collaborations between pairs of authors. The overlay coloring encodes the average publication year associated with each author, thereby revealing the temporal layering of collaboration communities and distinguishing historically established groups from more recent or emerging collaboration patterns.

The network exhibits a multi-component structure dominated by several relatively compact collaboration groups rather than a single, highly connected core. This pattern is consistent with Granta tools being used across diverse application domains and institutional contexts. A historically anchored cluster is centered on foundational materials-selection and materials engineering contributors (e.g., \textit{Ashby, M.~F.} with 26 documents and total link strength 15; \textit{Cebon, D.} with 10 documents; and \textit{Br\'echet, Y.} with 8 documents), with early average publication years (approximately 2003--2008), reflecting long-standing methodological and educational roots.

More recent thematic communities appear as separate collaboration groups with later overlay years. For example, a dense cluster around \textit{Bontempi, E.} (12 documents; total link strength 27) and close collaborators (e.g., \textit{Depero, L.~E.}, \textit{Federici, S.}, \textit{Zanoletti, A.}) is characterized by average publication years around 2019--2021, consistent with a more contemporary wave of applied studies. Additional mid-period clusters (around 2016--2018) include recurring co-author teams such as \textit{Fredriksson, C.} (18 documents) and collaborators, as well as process/manufacturing-oriented groups (e.g., \textit{Ingrao, G.}--\textit{Priarone, P.~C.}--\textit{Settineri, L.}).

Finally, a small number of peripheral nodes show very recent average publication years (e.g., 2023--2024), indicating emerging or newly active author groups with limited connectivity to earlier clusters. Overall, the author network highlights Granta product usage as a distributed ecosystem: a persistent historical backbone linked to materials selection and education, complemented by multiple more recent applied clusters that evolve in parallel across domains.

\FloatBarrier
\subsection{Keyword and abstract term mapping}
\label{subsec:kw_map}

To provide an initial qualitative overview of the corpus content, exploratory text statistics were computed from publication titles and keywords after basic preprocessing (lowercasing and stop-word removal). The most frequent terms are dominated by \textit{material} and \textit{design}, confirming that the corpus is primarily centered on materials-oriented engineering and design workflows. More specifically, the top-ranked expressions include \textit{material} (3.83\%), \textit{design} (2.74\%), and \textit{material selection} (1.12\%), directly reflecting the central role of materials choice and trade-off analysis in Granta-supported activities.

Beyond these core concepts, the prominence of terms such as \textit{sustainability} (1.07\%), \textit{life cycle} (0.97\%), and \textit{analysis} (1.03\%) indicates a strong coupling between materials data usage and environmental or performance-driven evaluation frameworks. The presence of \textit{additive manufacturing} (0.86\%), \textit{composite} (0.81\%), and \textit{manufacturing} (0.69\%) further highlights the importance of advanced materials and processes in the application landscape, while the term \textit{data} (0.73\%) reflects the data-centric nature of the reported workflows.

\begin{figure}[h]
\centering
\includegraphics[width=0.95\textwidth]{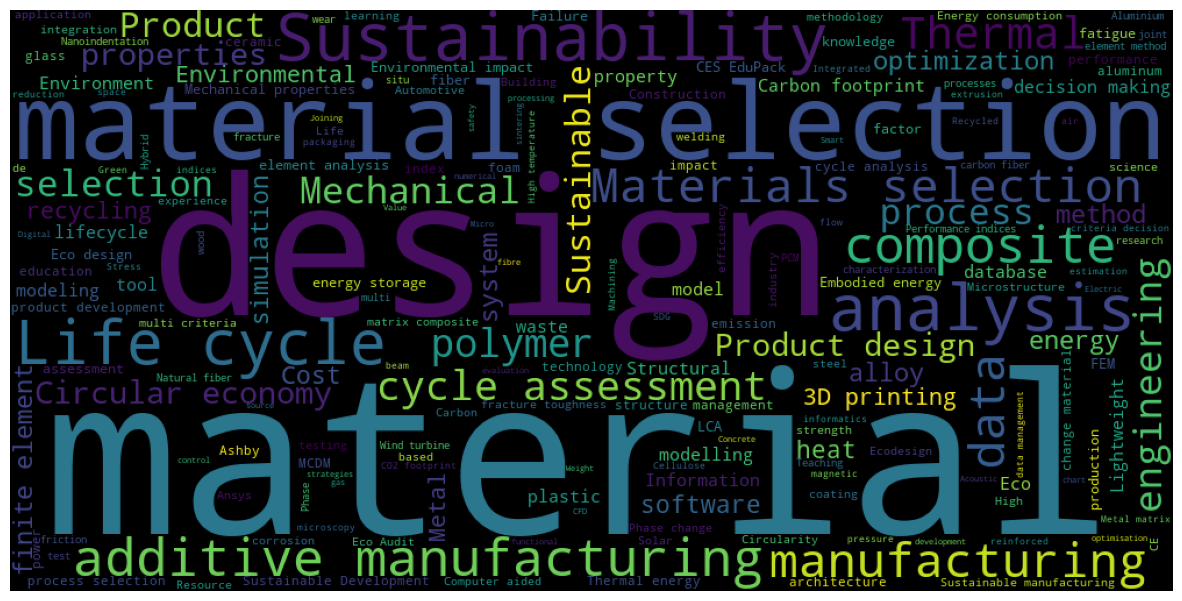}
\caption{Word cloud generated from publication keywords in the curated corpus, highlighting the most frequent technical terms and application themes associated with Ansys Granta product usage.}
\label{fig:WordCloud_Keyword}
\end{figure}

Complementary text-based summaries were also generated. In particular, keyword-based term prominence was visualized using a word cloud (Fig.~\ref{fig:WordCloud_Keyword}) to provide a compact overview of recurring technical terms and application themes. This representation is used as an exploratory complement to the structured domain and application-field classifications encoded in the curated metadata, and serves primarily as a qualitative support to the quantitative bibliometric indicators discussed above.

\FloatBarrier

\begin{figure}[h]
\centering
\includegraphics[width=0.95\textwidth]{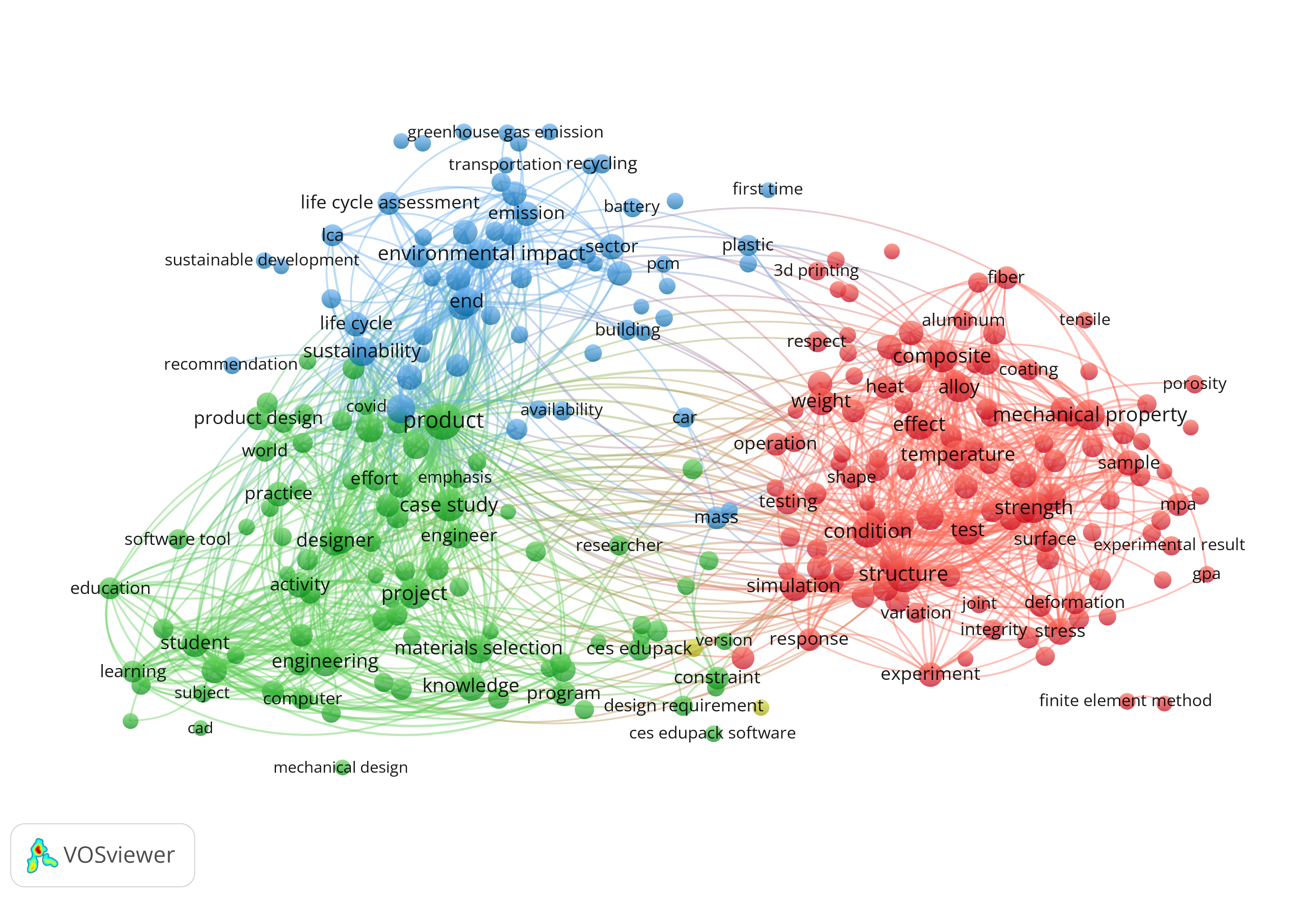}
\caption{VOSviewer term co-occurrence map generated from publication keywords and abstracts in the curated corpus. Node size reflects term frequency and link strength reflects co-occurrence intensity. Colors indicate clusters, revealing three main thematic groups: design and product development, materials science and engineering, and a smaller sustainability- and eco-properties-oriented cluster.}
\label{fig:KW_Map}
\end{figure}

To complement the descriptive indicators and field-based classification, a term co-occurrence map was generated from publication keywords and abstracts using VOSviewer. After standard text preprocessing (e.g.\ normalization and removal of non-informative terms), the resulting network was clustered based on term co-occurrence, yielding the thematic structure shown in Fig.~\ref{fig:KW_Map}. In this representation, nodes correspond to terms and links reflect their tendency to appear together within the same documents, while the clustering highlights groups of terms that define coherent application themes.

Three main thematic clusters can be identified. The first and largest cluster is primarily oriented toward \textit{design and product development} and teaching aspect, covering materials selection in early-stage engineering, decision-support, and design-oriented workflows. A second major cluster is more strongly anchored in \textit{materials science and engineering}, reflecting studies focused on material behavior, characterization, processing, and materials-specific research applications. In addition, a smaller and more peripheral cluster is centered on \textit{sustainability and eco-properties}, including terms related to environmental impact, life-cycle assessment, eco-design, and sustainable materials.

Taken together, these clusters confirm the multi-faceted positioning of Granta usage in the literature: as a design-enabling decision layer, as a materials engineering data and analysis backbone, and increasingly as a support tool for sustainability-driven and environmentally conscious engineering practices.

\FloatBarrier

\section{Results: Mapping Ansys Granta Product Usage}
\label{sec:results}

To establish a quantitative overview of product adoption, we first examine how Ansys Granta products are referenced and applied across the reviewed publications. Approximately three quarters of the papers cite Granta tools primarily as sources of curated materials data or databases. Materials selection emerges as the predominant application (about 40\% of studies), whereas process selection is addressed far less frequently (6\%). Additionally, 34\% of the publications leverage the software’s visualization and charting capabilities, and 10\% explicitly report the use of the Eco Audit tool.

\subsection{Temporal evolution of product usage}
Publication volume and product mentions were tracked over time to identify adoption dynamics and potential external drivers.
A pronounced increase in \textit{Granta EduPack} mentions is observed around 2020, which is consistent with the rapid shift to remote and hybrid teaching during the COVID-19 pandemic, and the resulting demand for digital pedagogical resources, software-supported materials education, and simulation-enabled learning. This trend aligns with the discussion of post-pandemic training challenges and the role of software tools for materials education reported in the SF2M white paper chapter on post-pandemic education \cite{Dwek2021SF2MFormation}. The comparatively lower number of publications observed at the end of 2025 should be interpreted with caution, as the document collection stops in early 2025 and therefore does not reflect a full publication year.
In addition, the marked increase in \textit{Granta MI} usage observed between approximately 2015 and 2020 may be attributed, at least in part, to the strong involvement of Granta Design Ltd. and subsequently Ansys Inc. in multiple European Union–funded research projects, which promoted the adoption of materials data management platforms within collaborative research and industrial workflows.

\begin{figure}[h]
\centering
\includegraphics[width=0.95\textwidth]{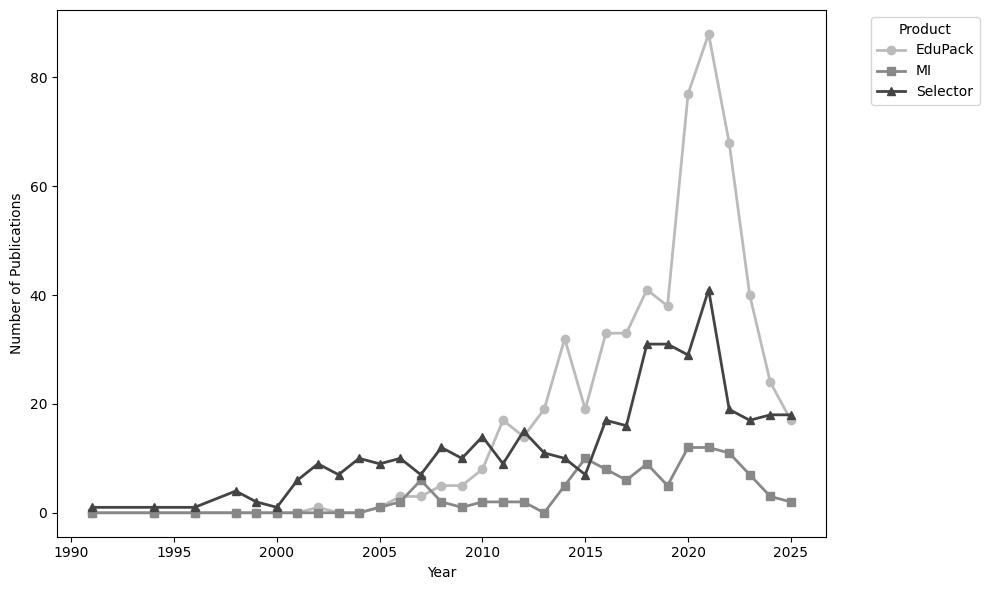}
\caption{Temporal evolution of Ansys Granta product usage based on yearly mentions in the analyzed corpus. The peak in \textit{Granta EduPack} around 2020 is consistent with pandemic-driven remote teaching needs and increased reliance on digital tools for materials education.}
\label{fig:TemporalEvolutionProducts}
\end{figure}

\FloatBarrier

\subsection{Key contributors and institutions}
\label{subsec:key_institutions}

\begin{figure}[h]
\centering
\includegraphics[width=0.95\textwidth]{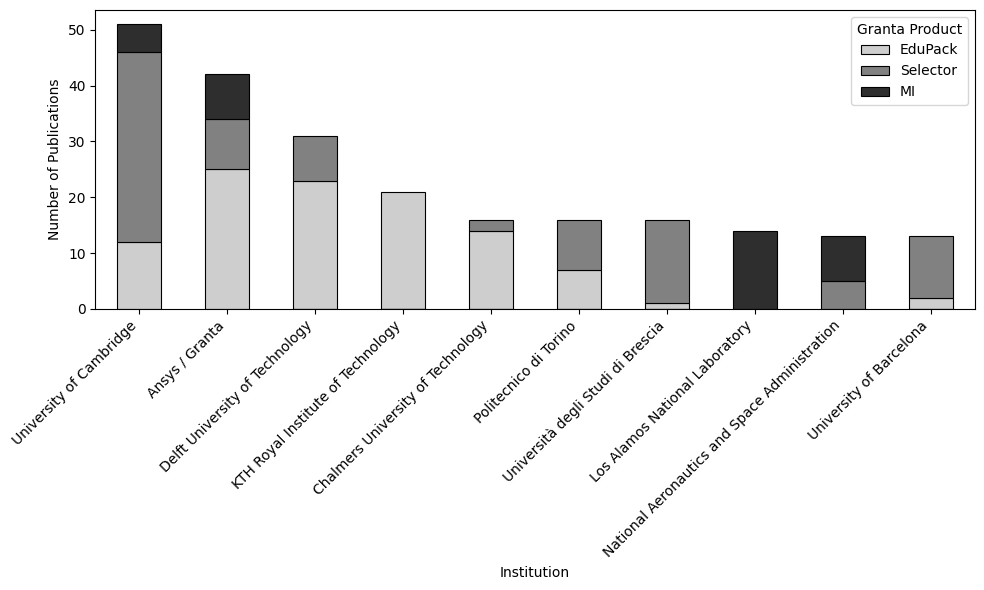}
\caption{Top 10 institutions ranked by the number of publications referencing Ansys Granta products, with counts shown as a stacked distribution by product (\textit{EduPack}, \textit{Selector}, and \textit{MI}). The ranking includes leading technical universities as well as industrial and governmental organizations, illustrating both academic and applied engagement and differences in product usage across institutions.}
\label{fig:Best10Univ}
\end{figure}

This subsection examines the institutions most actively publishing works that report substantive use of Ansys Granta products, providing insight into the academic, industrial, and governmental ecosystems that drive adoption and dissemination. In total, more than 330 distinct institutions were identified across the curated corpus, illustrating the broad organizational diversity of the Granta user base and its penetration across research, education, and applied engineering contexts. For consistency and to avoid double counting, institutional statistics were computed using the \emph{first-listed author affiliation only}. In other words, each publication was attributed to the primary affiliation of the first author (typically corresponding to the main contributing or coordinating institution). Co-author affiliations were not included in the counting process. This choice ensures a unique assignment of each record to a single institution and provides a clear and reproducible ranking, although it may underestimate the broader collaborative involvement of partner organizations. Figure~\ref{fig:Best10Univ} presents the ten institutions with the highest number of Granta-related publications. Counts are displayed as stacked bars by principal product (\textit{EduPack}, \textit{Selector}, and \textit{MI}), enabling direct comparison of product-specific usage profiles and revealing differences in institutional focus, ranging from education-oriented usage to enterprise materials data management and simulation-driven engineering workflows.

\FloatBarrier

The list is dominated by major technical universities, including the University of Cambridge, Delft University of Technology, KTH Royal Institute of Technology, Chalmers University of Technology, Politecnico di Torino, Università degli Studi di Brescia, and the University of Barcelona. Their contributions are largely associated with EduPack and Selector, reflecting the widespread use of these tools in education, materials selection, and design-oriented research.

Industrial and governmental organizations are also strongly represented. Ansys Inc./Granta Design Ltd. shows significant activity across all products, while Los Alamos National Laboratory and the National Aeronautics and Space Administration exhibit a comparatively stronger use of Granta MI, consistent with enterprise-scale materials information management and simulation-driven engineering applications.

In the specific case of the University of Cambridge, its leading position is consistent with the historical origin of Granta Design Ltd. in Cambridge, which likely fostered early adoption, close collaboration, and sustained visibility of Granta tools within the local research ecosystem.

\subsection{Disciplinary and application domains}

This subsection analyzes the distribution of Ansys Granta usage across scientific and engineering fields and materials families, based on the curated metadata and bibliometric indicators described above. The analysis aims to characterize the disciplinary positioning of Granta technologies and to identify dominant and secondary domains of application within the literature. To improve comparability with established bibliometric and research and development (R\&D) reporting practices, the application fields were mapped to the Organisation for Economic Co-operation and Development (OECD) \emph{Fields of Science and Technology} (FOS, 2007) taxonomy at the level of second-order categories (e.g., Mathematics; Computer and information sciences; Mechanical engineering; Materials engineering) \cite{Grabowska_2019}.

\begin{figure}[h]
\centering
\includegraphics[width=0.95\textwidth]{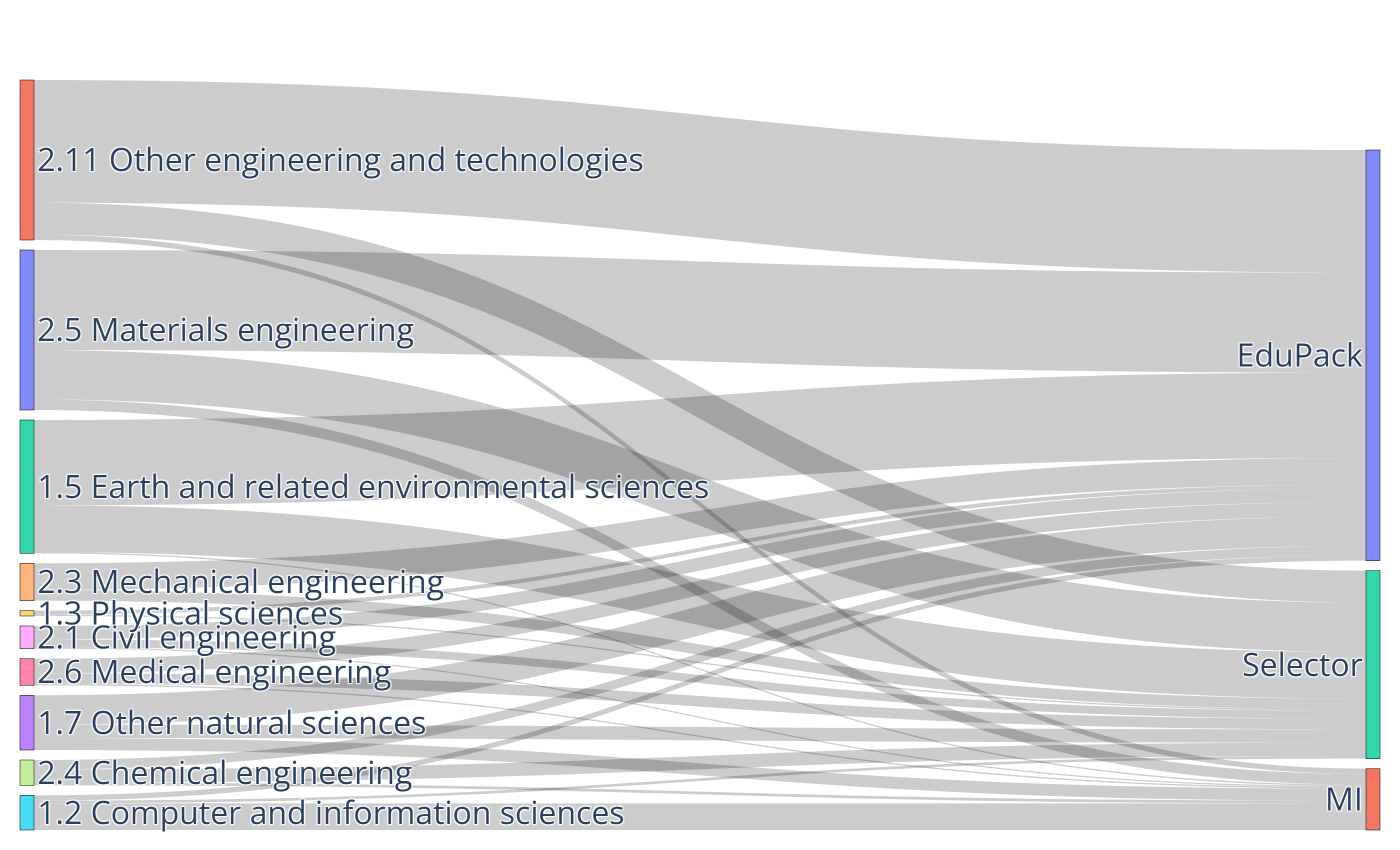}
\caption{Sankey diagram illustrating the relationships between OECD FOS (2007) application fields and Ansys Granta products. The width of each flow is proportional to the number of documents associating a given OECD FOS field with a specific Granta product.}
\label{fig:SankeyFieldProducts}
\end{figure}

Figure~\ref{fig:SankeyFieldProducts} presents a Sankey diagram illustrating the relationships between OECD FOS (2007) application fields and Ansys Granta products. Each record was assigned to a single OECD FOS second-order category based on cleaned and harmonized application field labels. The width of each flow is proportional to the number of documents associating a given disciplinary field with a specific Granta product. This representation highlights the strong concentration of Granta usage within engineering and materials-related fields, with particularly strong links to materials engineering, mechanical engineering, and related applied sciences.

\FloatBarrier

\begin{figure}[h]
\centering
\includegraphics[width=0.95\textwidth]{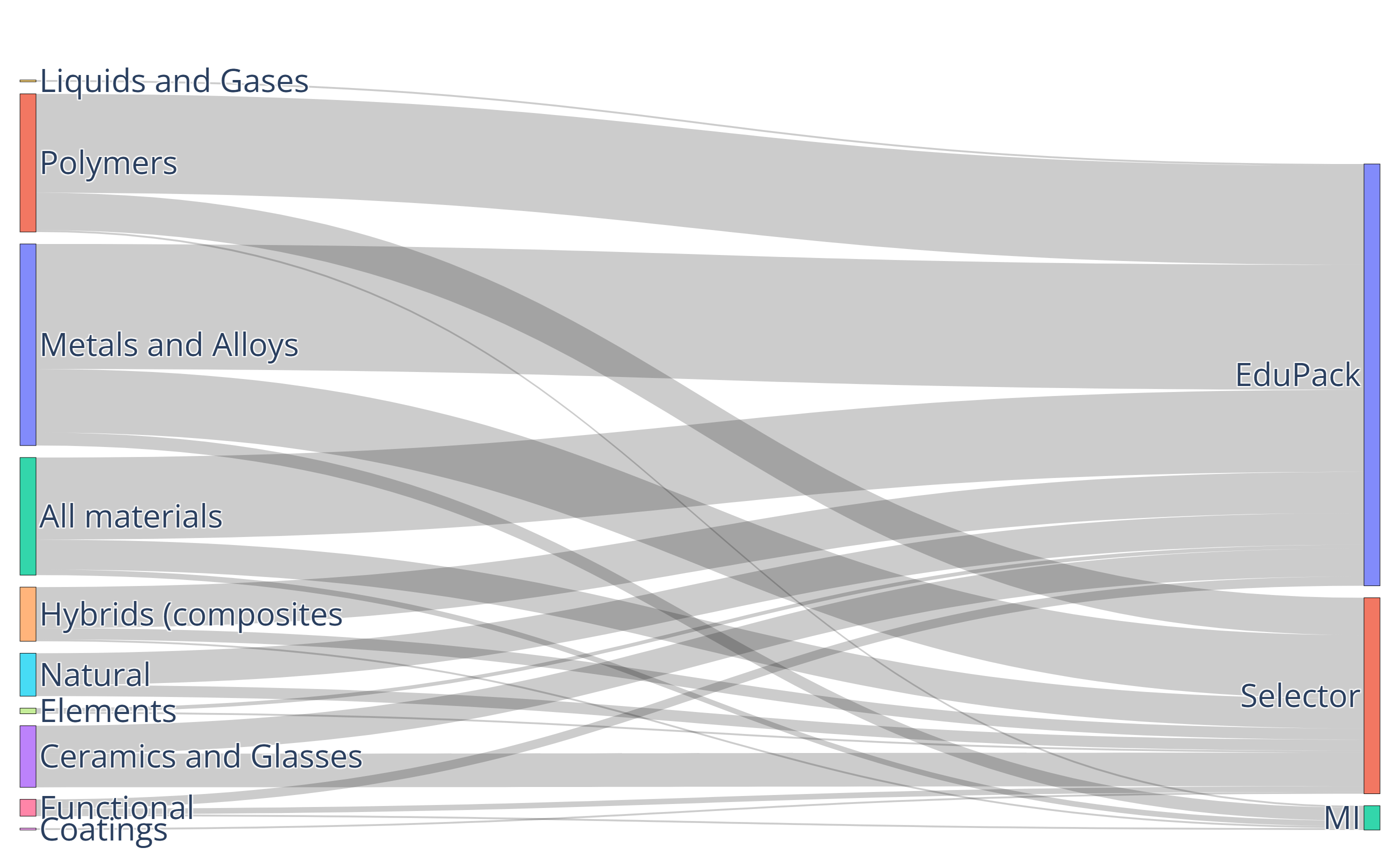}
\caption{Sankey plot showing how each material family distributes across Ansys Granta products. Link widths represent the relative percentage of publications within each material family associated with a given product, enabling proportional comparison across material classes.}
\label{fig:MaterialsVsProducts}
\end{figure}

Complementary to the disciplinary analysis, Figure~\ref{fig:MaterialsVsProducts} adopts a materials-centric perspective by examining how publications associated with each material family distribute across the different Ansys Granta products. The relationships are visualized using a Sankey plot, where link widths are normalized to represent relative proportions rather than absolute document counts. This normalization enables direct comparison of usage patterns across material families of different sizes. The results show that most material families exhibit a similar usage structure dominated by EduPack. For major categories such as \emph{Metals and Alloys}, \emph{Polymers}, and \emph{Ceramics and Glasses}, the majority of publications are connected to EduPack, indicating that educational, exploratory, and early-stage materials selection activities constitute the primary mode of interaction with Granta tools across materials domains. Selector contributes a secondary but consistent share across these same families, suggesting its role in comparative screening, trade-off analysis, and engineering decision-making once candidate materials have been identified. Its presence appears proportionally stable rather than concentrated in any single material class. In contrast, MI accounts for only a small fraction of usage within most families. Its comparatively limited share, coupled with a more even distribution across broad or generic material categories, points toward specialized adoption in enterprise-scale materials data management workflows rather than routine research or teaching activities. Overall, the materials-oriented view confirms that, regardless of the specific material class considered, usage patterns are structurally similar.

\FloatBarrier

\subsection{Combination with CAD/FEM and CAE software}

Beyond their standalone usage, Ansys Granta products are in a significant subset of cases combined with external CAD, CAE, and FEM software tools, reflecting their integration into broader digital engineering workflows. In the curated corpus, 203 publications (approximately 20\% of the total dataset) explicitly report a combined use of Granta products with at least one CAD/CAE or FEM environment. Although such coupling is not systematic, it represents a substantial and structurally meaningful usage pattern.

\begin{figure}[h]
\centering
\includegraphics[width=0.75\textwidth]{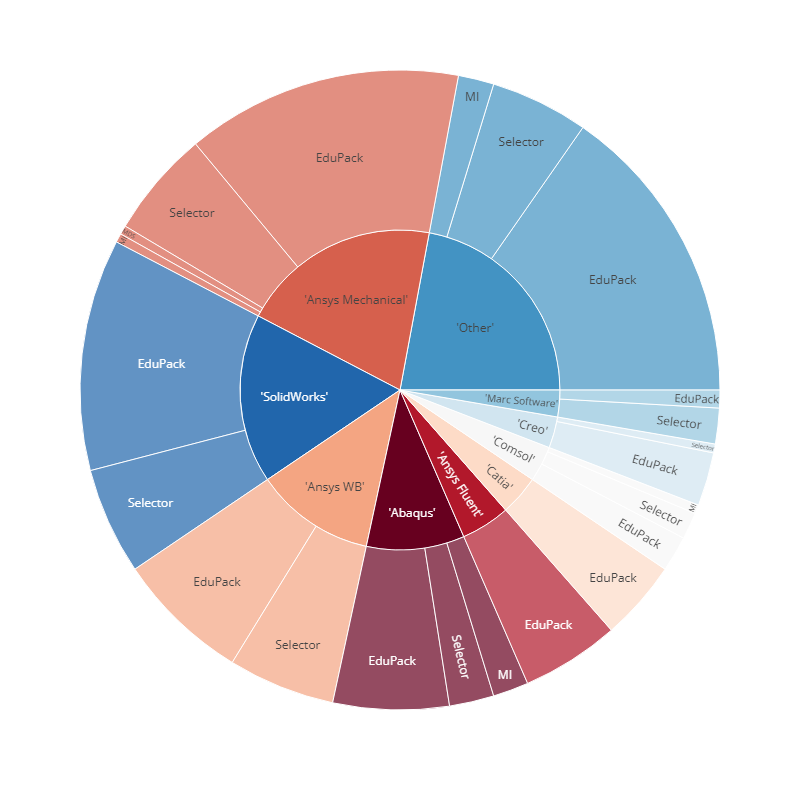}
\caption{Distribution of CAD, FEM, and CAE software tools reported in combination with Ansys Granta products in the curated corpus. The chart highlights the strong coupling between Granta tools and major simulation and design environments, particularly within the Ansys ecosystem, as well as with widely used external CAD and multiphysics platforms.}
\label{fig:CombinationCADFEM}
\end{figure}

Figure~\ref{fig:CombinationCADFEM} summarizes the main software environments reported alongside Granta products in these combined-use cases. The most frequently associated environment is the generic or heterogeneous category grouped as \textit{Other} (46 occurrences, $\sim$21\%), closely followed by \textit{Ansys Mechanical} (45 occurrences, $\sim$21\%), and the widely adopted CAD platform \textit{SolidWorks} (38 occurrences, $\sim$17\%). This ranking highlights both the strong affinity between Granta tools and the Ansys simulation ecosystem and their broad integration within mainstream mechanical design workflows.

The continued presence of \textit{Ansys Workbench} and established external solvers such as \textit{Abaqus}, \textit{COMSOL}, \textit{Marc}, and \textit{Ansys Fluent} further emphasizes the role of Granta products as upstream providers of validated materials and process data to numerical simulation environments. Similarly, the co-occurrence with commercial CAD systems including \textit{CATIA} and \textit{Creo} illustrates the use of Granta tools in early-stage design and materials-driven decision-making. Taken together, these patterns confirm that while many reported Granta usages remain focused on materials selection, education, and data management in isolation, a non-negligible fraction of the literature positions Granta products as a materials knowledge layer embedded within heterogeneous CAD/CAE ecosystems, particularly in ICME-oriented and simulation-driven engineering practices.

\FloatBarrier

\subsection{Thematic structure and emerging trends}
\label{subsec:trends}
\begin{figure}[h]
\centering
\includegraphics[width=0.95\textwidth]{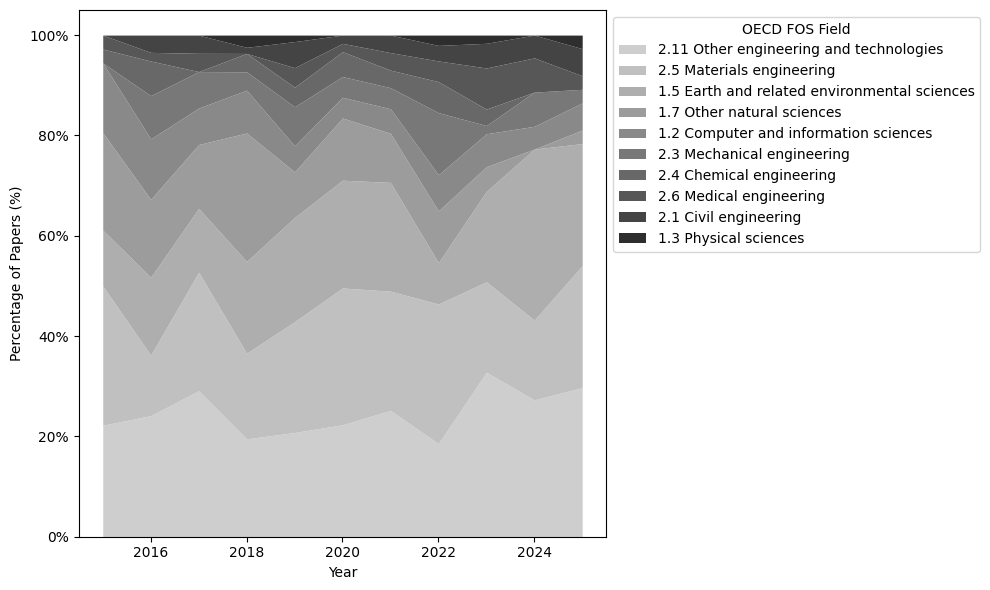}
\caption{Temporal evolution of OECD FOS (2007) fields in the curated corpus (2015--2025). Counts represent the number of publications assigned to each field per year. The apparent decline in 2025 reflects incomplete coverage due to the dataset collection stopping in early 2025.}
\label{fig:FieldVsTime}
\end{figure}

To characterize recent thematic dynamics, the OECD FOS field assignment was analyzed as a function of publication year over the last decade (2015--2025). As shown in Fig.~\ref{fig:FieldVsTime}, the distributions are expressed in relative values (percentage of articles per year), allowing comparison of thematic structures independently of annual publication volume.

The corpus is consistently dominated by \textit{Materials engineering} and \textit{Other engineering and technologies}, reflecting the central positioning of Granta tools at the intersection of materials data, selection methodologies, and engineering practice. A pronounced multi-field expansion is observed in 2020--2021, with simultaneous increases in \textit{Materials engineering}, \textit{Other engineering and technologies}, and \textit{Earth and related environmental sciences}, consistent with the broader growth of data-driven materials research and sustainability-oriented studies during this period.

Beyond these dominant categories, several fields exhibit smaller but noteworthy signals. \textit{Medical engineering} shows a visible rise after 2019 with sustained contributions through 2021--2023, suggesting a growing adoption of Granta-supported methodologies in biomedical and healthcare-related materials applications. \textit{Mechanical engineering} remains a secondary contributor overall but displays a local maximum around 2022, in line with increased reporting of simulation- and design-driven workflows. \textit{Computer and information sciences} remains present at a comparatively lower level throughout the period, indicating that while data-centric aspects are increasingly visible, the corpus remains primarily anchored in applied engineering and materials domains rather than in computer science venues.

Finally, the year 2025 should be interpreted with caution, as the database collection stops in early 2025 and therefore does not represent a complete publication year.

\FloatBarrier

\section{Conclusion}
\label{sec:conclusion}

This paper presented a curated, software-centric literature database implemented in Ansys Granta MI Enterprise to monitor more than three decades of Ansys Granta product usage across research, industry, and engineering education. By combining semi-automated bibliographic ingestion (e.g., DOI and citation-file parsing) with expert curation of usage-oriented metadata, the proposed approach enables a traceable, queryable, and extensible representation of how Granta tools are operationally employed in the scientific and technical literature.

Based on a curated corpus of more than 1{,}100 records spanning 1990--2025, the analyses reveal distinct temporal, institutional, and disciplinary patterns across products. The results show a strong dominance of Granta EduPack in the literature, consistent with its broad adoption in materials education and early-stage design studies, while Granta Selector exhibits a more specialized presence aligned with comparative materials screening workflows. Granta MI appears less frequently in academic publications but is associated with contexts consistent with enterprise-scale materials data management. At the workflow level, approximately 20\% of the curated publications explicitly report coupling Granta tools with external CAD/CAE/FEM environments, highlighting a non-negligible integration pattern in simulation-driven and ICME-oriented practices.

Beyond descriptive indicators, the database architecture and relational links between documents, authors, institutions, products, and domains provide a practical foundation for reproducible analytics, including network representations and interactive dashboards. As such, the proposed framework supports multiple technically driven use cases, from rapid retrieval of validated case studies and reproducible literature reviews to technology scouting and early detection of emerging themes relevant to product strategy.

Several limitations should be noted. The database is non-exhaustive and depends on the discoverability of software-use evidence in publications, which remains constrained by inconsistent citation practices and heterogeneous document types. Future work will focus on extending coverage through improved semi-automated discovery, strengthening normalization of entities (authors, institutions, and software names), and enriching usage descriptors to better capture product versions, integration pathways, and levels of methodological dependence. More broadly, the methodology introduced here is transferable to other engineering software ecosystems where systematic monitoring of software adoption and application domains is required.

\section*{Authors Contributions}

D.~Mercier conceived the study, designed the database schema and curation protocol, populated and curated the reference database, developed the data extraction and analysis scripts, performed the bibliometric and visualization analyses, and wrote the manuscript.

\section*{Acknowledgements}

The author gratefully acknowledges Jean-Marc Lucatelli, Victor Étique, and Mickaël Capelli for their technical support and contributions to the design and implementation of the database architecture. The author also thanks the Ansys Academic Team (mostly Nicolas Martin, When Zhao, Kaitlin Tyler) for their assistance in identifying new reference papers and supporting the archiving process within the database, as well as Davide Di Stefano and Pascal Salzbrenner for their rich and insightful discussions, which helped shape and refine this work.

\section*{Declaration of Generative AI and AI-Assisted Technologies in the Manuscript Preparation}

During the preparation of this manuscript, the author(s) used Microsoft Copilot for grammar correction, syntax refinement, and improvements in clarity. After using this tool, the author(s) carefully reviewed and edited the content as needed and take full responsibility for the final version of the published article.

\bibliographystyle{unsrtnat}
\bibliography{references}

\end{document}